\documentclass{edp-conf}
\usepackage{graphicx}
\begin{document}

\TitreGlobal{SF2A 2003}

\title{Modulation of the X-ray Fluxes by the Accretion-Ejection Instability}
\author{Varni\`ere, P.}\address{University of Rochester, Department of Physics \& Astronomy}
\author{Muno, M.}\address{MIT}
\author{Tagger, M.}\address{SAp, CEA/Saclay}
\author{Frank, A}\address{University of Rochester, Department of Physics \& Astronomy}
\runningtitle{Modulation of the X-ray Fluxes by the AEI}
\setcounter{page}{237}
\index{Varni\`ere, P.}
\index{Muno, M.}
\index{Tagger, M.}
\index{Frank, A.}

\maketitle
\begin{abstract}
The Accretion-Ejection Instability (AEI) has been proposed to explain the low frequency Quasi-Periodic 
Oscillation (QPO) observed in low-mass X-Ray Binaries. Its frequency, typically a fraction of the Keplerian 
frequency at the disk inner radius, is in the right range indicated by observations. With numerical 
simulation we will show how this instability is able to produce a modulation of the X-ray flux and what are 
the characteristic of this modulation. More simulations are required, especially 3D MHD simulations. 
We will briefly present a new code in development: AstroBear which will allow us to create synthetic spectra. 
\end{abstract}
%
\section{Introduction: a brief presentation of the AEI and relation to observation}
  

      The Accretion-Ejection Instability [\cite{TP99}] is a spiral instability similar to the galactic spiral
      but driven by magnetic stress instead of self-gravity. This instability affects the 
      inner region of the disk when the plasma $\beta = 8\pi p/B^2$ is of the order of one, 
      i.e. there is equipartition between the gas and magnetic pressure. It forms a
      quasi-steady spiral pattern rotating in the disk at a frequency of the order of
      a few times the orbital frequency at the inner edge of the disk.
      This spiral density wave is coupled with a Rossby vortex it creates at the corotation radii 
      (corotation between the spiral wave and the gas in the diks). This coupling permits
      the energy and angular momentum to be stored at the corotation radius  allowing 
      accretion to proceed in the inner part of the disk. Contrary to MRI based accretion models
      the AEI does not heat-up the disk as the energy is transported by waves and not deposited 
      locally.

      The Rossby vortex twists the foot points of the field liness. In the presence of a low density
      corona this torsion will propagate as an Alfven wave transporting  energy and angular
      momentum store in the vortex. This will put energy into the corona where it might power
      a wind or a jet [\cite{VT02}].
\newline

      In order to apply the AEI to the phenomena occuring in microquasars we have two observables that
      we used: the presence of jet; thecharacteristic of the Quasi-Periodic Oscillation (QPO).
      We already have compared some of the properties of the AEI with observations, mainly the
      relation between the inner radius of the disk from  spectral fit  and the
      QPO frequency [\cite{R02}], [\cite{V02}]. This  leads us to try to compare more  QPO
      characteristics with the observations.
      
\section{AEI and QPO}

\begin{figure}[h]
\centering
\includegraphics[width=5cm]{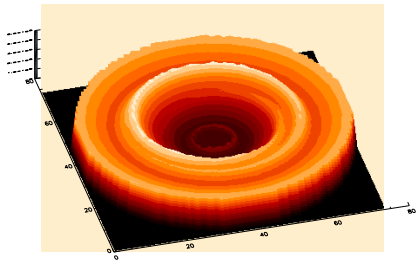}
\caption{Snapshot of the inner disk, the heigh represent the local thickness of the disk
      and the color represent the temperature (lighter color meaning higher temperature)}
\label{fig:hot_point}
\end{figure}

     In order to compare the AEI to observations we  aim to produce synthetic spectra
     from numerical simulation. We used the code presented in [\cite{CT01}] and add an
     energy equation to compute the heating of the disk at spiral shocks. The idea was to use
     the spiral shock to heat the disk and create a thicknenning along the spiral.
     Using an hydrostatic approxmation to compute the local thickness of the disk we get
     the inner region of the accretion disk appears as in figure \ref{fig:hot_point}. We 
     see that along the spiral arm the disk gets thicker (the $z$ coordinates)  and hotter (the color scale)
     than other locations in the disk.

\begin{figure}[h]
\centering
\begin{tabular}{cc}
\includegraphics[width=5cm]{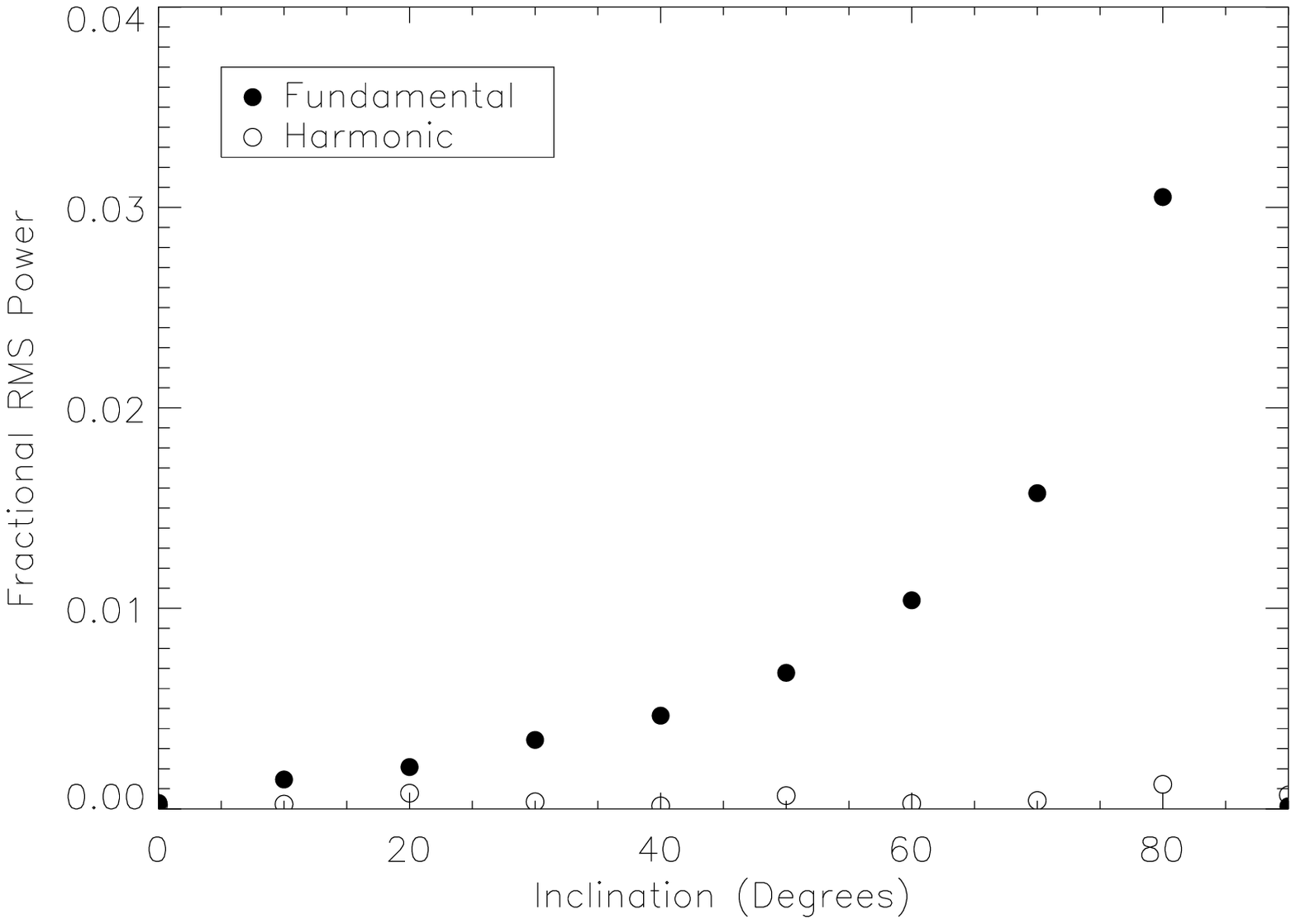}  &
\includegraphics[width=5cm]{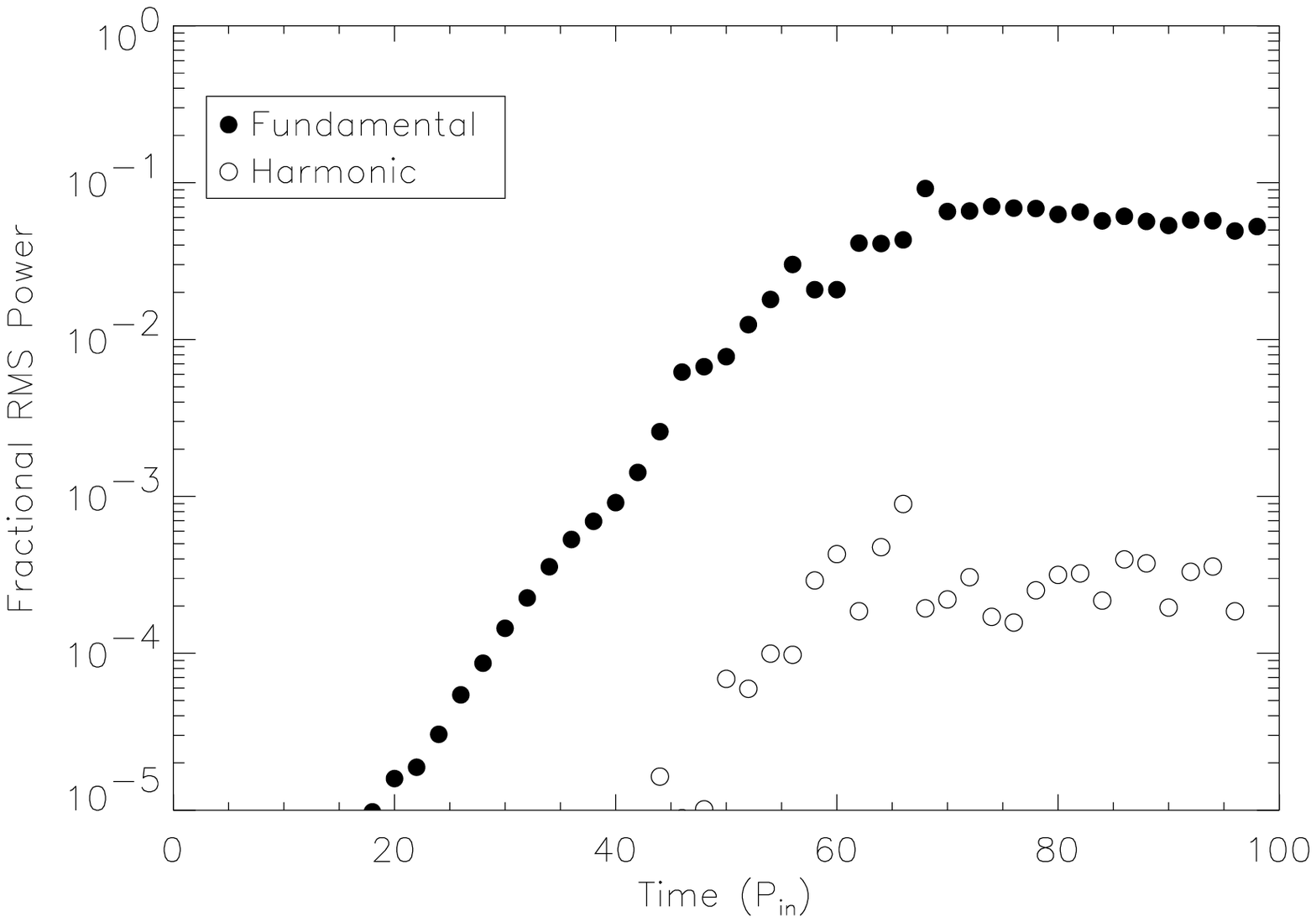}
\end{tabular}
\caption{left: maximun rms amplitude obtain for different inclination of the system.
right: evolution of the rms amplitude of the modulation as function of time.}
\label{fig:QPO}
\end{figure}

     In collaboration with M.Muno we have computed the X-ray flux coming from such an accretion 
     disk. We obtain a modulation of the observed flux. This modulation is coming from 
     geometrical effect related to the inclination angle of the source such as seen on the
     left graph of figure \ref{fig:QPO}. 

     We see that the rms amplitude ($\sim 5\%$) is too small  to explain the 
     observations (as much as $20\%$ sometimes). Several phenomena could increase the 
     observed rms amplitude of the modulation. Indeed, the simulation did not take
     into account relativistic effects.      
     Another effect is the fact that the disk
     thickness is computed as posteriori and not evolve in the simulation. The
     hydrostatic equilibrium is a good ``first approximation'' and allows
     us to prove that the AEI is able to modulate the X-ray fluxes but in order
     to compare with observation we need simulations of the disk taking into account
     the three dimensions.

\section{A New $3$D MHD AMR code: AstroBear}

     Numerical simulation is a tool which allows us to study phenomena and also to
     compare theory with observation by the mean of synthetic observation/spectra in
     their non-linear behaviour. 
     If we want to study in more detail the accretion and ejection we need to have a 
     $3$D MHD code. 
     Including Adaptativ Mesh Refinement (AMR) in the code will allow us to have
     a good resolution with less numerical cost. Using AMR is not always useful depending
     on the phenomena study, i.e. AMR is not interesting for turbulence but it will be really
     useful to study jet's knots or spiral waves. 
\newline


    We are working on such a $3$D AMR MHD code using a Godunov-type methods for the base scheme.
    This method  projects the solution of the eigenfunction of the Riemann problem 
    associated.

    Astro-Bear use a package called BEARCLAW.
    BEARCLAW is  a general purpose software package for solving time dependent
    partial differential equations (automatic adaptive mesh refinement, parallel execution).
    AstroBear is a module meant for astrophysical purpose, especially the accretion-ejection
    phenomena. At the moment the hydrodynamics module fully operationnal in $3$D and we
    are testing the MHD.
\newline


    We have done the standard $1$D MHD test to check the ability of the code to
    resolve each waves. 
    When going to $2$D one needs to take care of the divergence of the magnetic field.
    Several method exist currently  in AstroBear we utilize three of them, the $8$th-waves, the GLM and
    the projection method. Our interest is
    to find the one that is the most adapted depending on the problem we want to solve.
    We will also add the choice between several Riemann solver, each of
    them having different weakness and strength.

    In figure \ref{fig:test_2D} show two standard $2$D tests. On the left is a
    cloud-shock interaction where a shock wave propagates through a media having a high 
    density cloud in it. To do this test we used the $8$th-waves method for the
    divergence constraint. This method does not ``clean'' the divergence mkdir at each step 
    but advect it with the flow. It does not required a new step but add the non-vanishing
    divergence term in the source step. It is the fastest method and well adapted.
\begin{figure}[h]
\centering
\begin{tabular}{cc}
\end{tabular}
\caption{2D test of the MHD solver with two differents method for the divergence.
left: a cloud-shock interaction using the 8-waves method. 
right:the Orzag \& Tang vortex using the projection method.}
\label{fig:test_2D}
\end{figure}
    The figure on the right is the Orzag \& Tang vortex. This time we have used the
    projection method in order to ``clean'' the divergence at each time step. This
    method requires us to solve a Poisson equation and cost a lot in term of efficiency.

\section{conclusion}

    Using numerical simulation we gave the proof of principle that the AEI is able
    to modulate the X-ray flux and we aim to continue this study by creating 
    synthetic spectra which can be directly compare with observation.
    
    In order to do this we need a code able to fully simulate an accretion disk.
    We have done the first step toward this goal by developing and testing the
    MHD module of AstroBear.


\end{document}